\pdfoutput=1

\documentclass[
  onecolumn,  
  nofootinbib,showpacs,superscriptaddress,preprintnumbers]{revtex4-2}

\usepackage[utf8]{inputenc}
\usepackage{graphicx, color}

\usepackage[svgnames,psnames]{xcolor}
\usepackage[colorlinks,citecolor=DarkGreen,linkcolor=FireBrick,linktocpage]{hyperref}

\usepackage{dcolumn}
\usepackage{bm}
\usepackage{amsmath}
\usepackage{amsthm}
\usepackage{ascmac}
\usepackage{xparse} 
\usepackage{mathtools}  
\usepackage{blkarray} 
\usepackage{amscd,verbatim}
\usepackage{amssymb}
\usepackage[all]{xy}
\usepackage{tikz}
\usepackage{enumitem} 
\usepackage{tikz-cd}
\usetikzlibrary{matrix, calc, arrows}

\newcommand\p{\partial}

\definecolor{mydarkred}{RGB}{233,20,35}
\definecolor{mypurple}{RGB}{120, 35, 160}
\definecolor{mydarkpurple}{RGB}{128, 100, 162}
\definecolor{mybrown}{RGB}{255, 195, 0}
\definecolor{myaqua}{RGB}{29, 153, 168}
\definecolor{myblue}{RGB}{91, 129, 184}  
\definecolor{mygreen}{RGB}{155, 187, 89}  

\definecolor{mybrightblue}{RGB}{0, 140, 255}

\usepackage[normalem]{ulem}  

\renewcommand\sout{\bgroup \color{red} \ULdepth=-.5ex \ULset}

\newcommand{\be}{\begin{equation}}
\newcommand{\ee}{\end{equation}}

\newcommand{\bea}{\begin{eqnarray}}
\newcommand{\eea}{\end{eqnarray}}

\newcommand{\bsp}{\begin{split}}
\newcommand{\espl}{\end{split}}

\newcommand{\bpm}{\begin{pmatrix}}
\newcommand{\epm}{\end{pmatrix}}

\newcommand{\bra}[1]{\langle {#1} |}
\newcommand{\ket}[1]{| {#1} \rangle}

\begin{document}

\title{ Squeezing stationary distributions of stochastic chemical reaction systems }

\author{Yuji~Hirono}
\email{yuji.hirono@gmail.com} 

\affiliation{
Asia Pacific Center for Theoretical Physics, Pohang, Gyeongbuk, 37673, Korea
}
\affiliation{
Department of Physics, Pohang University of Science and Technology, Pohang, Gyeongbuk, 37673, Korea
}
\affiliation{
RIKEN iTHEMS, RIKEN, Wako 351-0198, Japan
}

\author{Ryo~Hanai} 
\email{rhanai09@gmail.com} 

\affiliation{
Asia Pacific Center for Theoretical Physics, Pohang, Gyeongbuk, 37673, Korea
}
\affiliation{
Department of Physics, Pohang University of Science and Technology, Pohang, Gyeongbuk, 37673, Korea
}

\date{\today} 

\begin{abstract}
Stochastic modeling of chemical reaction systems based on master equations 
has been an indispensable tool in physical sciences. 
In the long-time limit, the properties of these systems are characterized by stationary distributions of chemical master equations. 
In this paper, we describe a novel method for computing stationary distributions analytically, based on a parallel formalism between stochastic chemical reaction systems and second quantization.  
Anderson, Craciun, and Kurtz showed that, 
when the rate equation for a reaction network admits a complex-balanced steady-state solution, the corresponding stochastic reaction system has a stationary distribution of a product form of Poisson distributions. 
In a formulation of stochastic reaction systems using the language of second quantization initiated by Doi, product-form Poisson distributions correspond to coherent states. 
Pursuing this analogy further, we study the counterpart of squeezed states in stochastic reaction systems. 
Under the action of a squeeze operator, the time-evolution operator of the chemical master equation
is transformed, and the resulting system describes a different reaction network, which does not admit a complex-balanced steady state. 
A squeezed coherent state gives the stationary distribution of the transformed network, for which analytic expression is obtained. 
\end{abstract}

\maketitle

\tableofcontents

\section{Introduction}

Modeling of chemical reaction systems is important for describing various systems in chemistry, physics, and biology~\cite{van1992stochastic,gardiner1985handbook,anderson2015stochastic}. 
When the number of molecules is large enough, fluctuations can be ignored, and 
deterministic rate equations for species concentrations can be used to track the time evolution of reaction systems. 
In contrast, in a situation where the number of molecules is not so large, the effect of fluctuations becomes important, and stochastic modeling is necessary. 
Stochastic reaction systems are commonly described by continuous-time Markov chains, whose time evolution is governed by chemical master equations.

The properties of stochastic reaction systems in the long-time limit are characterized by their stationary distributions. 
Calculating stationary distributions analytically\footnote{
For monomolecular reaction networks, 
time-dependent solutions of master equations 
can be obtained~\cite{jahnke2007solving}, which are parametrized by the solution of rate equations. 
} is in general a difficult task because a chemical master equation is a collection of infinitely many coupled ordinary differential equations. 
However, for a certain class of chemical reaction systems, stationary distributions can be obtained analytically~\cite{anderson2010product}: When the deterministic counterpart of a stochastic reaction system has a complex-balanced steady-state solution, the stationary distribution of the stochastic system is of a product-Poisson form, whose parameter is given by the deterministic steady-state solution. 
This can be combined with the 
classic result by Feinberg~\cite{FEINBERG19872229,feinberg2019foundations} and Horn--Jackson~\cite{horn1972general} that, if a chemical reaction network is of a zero deficiency and weakly reversible, it has a unique steady-state solution in each stoichiometry compatibility class, and the solution is complex-balanced. 
Hence, when a reaction network satisfies the two topological conditions, vanishing deficiency and weak reversibility, 
its stationary distribution is given analytically. 
To extend the applicability of these results, 
one possible strategy is to transform one network,
which does not satisfy the two conditions, 
into another that satisfies them, 
while keeping the same chemical properties. 
The method of network translation~\cite{johnston2014translated,johnston2019computing}, in which reactions with a common stoichiometry are combined to obtain another network with desirable topological properties, has been utilized~\cite{hong2021derivation}
to compute stationary distributions analytically for networks with nonzero deficiency and without weak reversibility.

\begin{figure}[htb] 
\centering
\begin{tikzpicture} 

\node at (2, 0) { 
{ 
$
\begin{tikzcd}[scale cd=1.4,sep=large]
\Gamma: \,\, 
\emptyset \rar["\ell_1", shift left] 
& A \lar["\ell_2", shift left]
\end{tikzcd}
$
} 
};  

\node at (7.2, 0) { 
 \scalebox{1.4} { $\Gamma': $} 
};  

\node at (9, 0) { 
{ 
$
\begin{tikzcd}[scale cd=1.3,sep=large]
\emptyset 
\rar["k_1", shift left] 
\dar["k_3"] 
& A \lar["k_2", shift left]
\\
2 A & 
\end{tikzcd}
$
} 
};  

\draw[mybrightblue, -stealth, line width=4pt] 
(4.3,0) -- (6.1,0); 

\node at (5.2,-0.6){ \scalebox{1.1} {\color{mybrightblue} Squeezing }};

\node at (5.2,-1.2){ \scalebox{1.2} {\color{mybrightblue} $S(\xi)$ }};

\node at (1.4, -1.7) { 
\color{mypurple}
 \scalebox{1.2} {
  Deficiency: 
  }
};

\node at (2.9, -1.7) { 
  \scalebox{1.3} {
  \color{mypurple} $\delta = 0$
  }
};

\node at (2, -2.5) { 
  \scalebox{1.2} {
  \color{mydarkred} Weakly reversible
  }
};

\node at (8.8, -1.7) { 
  \scalebox{1.3} {
  \color{mypurple} $\delta = 1$
  }
};

\node at (8.8, -2.5) { 
  \scalebox{1.2} {
  \color{mydarkred} Not weakly reversible
  }
};

\node at (1, -3.5) { 
  \scalebox{1.2} {
  Stationary dist.: 
  $\ket{\Psi} = \sum_n P(n) \ket{n}$ 
  }
};

\node at (9.5, -3.5) { 
  \scalebox{1.2} {
  $\ket{\Psi'} = 
  {\color{mybrightblue} S(\xi) }\ket{\Psi}  = \sum_n P'(n) \ket{n}$ 
  }
};

\end{tikzpicture}
\caption{ Schematic of the squeezing procedure. 
On the action of squeeze operator, 
the reaction network is transformed from $\Gamma$ to $\Gamma'$.
The transformed system $\Gamma'$ 
is of a nonzero deficiency and not weakly reversible. 
In the quantum-mechanical formulation of stochastic chemical reaction systems, the probability distribution is represented by a vector 
($\ket{\Psi}=\sum_n P(n)$ for $\Gamma$ and $\ket{\Psi'}=\sum_n P'(n)$ for $\Gamma'$). 
Since the network $\Gamma$ has deficiency zero and is weakly reversible, the stationary distribution follows the Poisson distribution.  
In the language of quantum optics, this state $\ket{\Psi}$ corresponds to a coherent state. 
The stationary distribution of network $\Gamma'$ 
can be obtained by operating a squeeze operator $S(\xi)$ on $\ket{\Psi}$, and the probability distribution $P'(n)$ 
can be obtained analytically (see Eq.~\eqref{eq:eg1-prob-analytic} for this example).
}
\label{fig:eg1-schematic}
\end{figure}
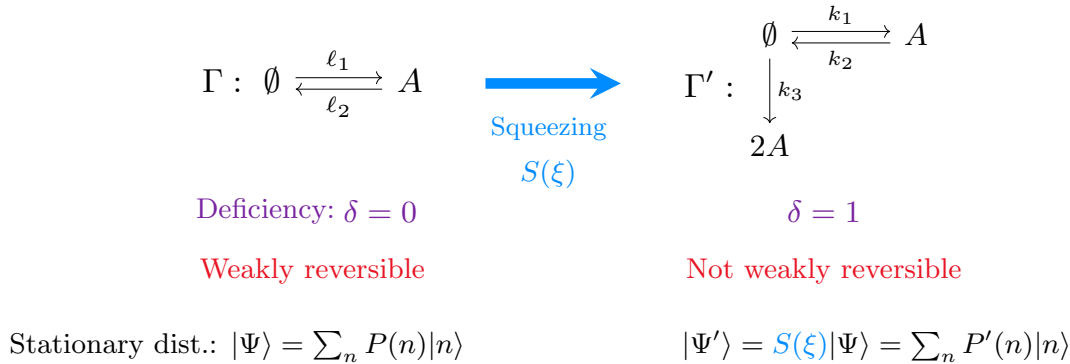

In this paper, we introduce a different kind of network transformation based on the parallel between stochastic reaction systems and quantum mechanics~\cite{Doi_1976, Doi_1976_2, baez2018quantum} (see Fig.~\ref{fig:eg1-schematic} for a general idea). 
There is a reformulation of stochastic reaction systems in which a probability distribution is represented by a vector spanned by the occupation-number basis.
In this formalism, a chemical master equation is written in the form analogous to the Schr\"{o}dinger equation.
Poissonian stationary distributions for complex-balanced systems correspond to coherent states in the context of quantum optics~\cite{Gardiner_Zoller}, that most-closely approximate classical states, saturating the uncertainty relation. 
We further pursue this analogy. 
Starting from coherent states, there are other states that can be reached by acting unitary operators. 
In particular, a common operation is squeezing, with which the uncertainty of a certain physical observable can be reduced at the cost of increasing the uncertainty of another observable. 
We find that we can perform squeeze operations on 
the Poissonian stationary distributions of complex-balanced systems. 
Under a squeeze transformation, the time-evolution operator is modified and it represents a different reaction network, which in general has a nonzero deficiency and is not weakly reversible. 
The stationary distribution of the transformed network is the counterpart of a squeezed coherent state in quantum mechanics, and its expression can be obtained analytically. 
Thus, the squeeze operation provides us with another way to analytically compute the stationary distributions of reaction networks which do not have complex-balanced steady-state solutions.

The remainder of the article is organized as follows.
In Sec.~\ref{sec:formulation}, 
we introduce the stochastic description of chemical reaction systems and its quantum-mechanical formulation using creation/annihilation operators. 
We also review the theorem by Anderson, Craciun, and Kurtz. 
In Sec.~\ref{sec:squeeze-single}, we introduce the squeeze transformation for a simple example. 
In Sec.~\ref{sec:two-mode-sq}, we derive the stationary distribution with finite correlations of different species. 
In Sec.~\ref{sec:nonlinear}, we discuss an example which involves nonlinear propensity functions. 
In Sec.~\ref{sec:generic}, we discuss the structural changes of generic reaction networks under squeezing transformations. 
Finally, we give a summary and further discussion in Sec.~\ref{sec:summary}.

\tikzcdset{scale cd/.style={every label/.append style={scale=#1},
    cells={nodes={scale=#1}}}}

\section{ Quantum-mechanical formulation of stochastic chemical reaction systems }\label{sec:formulation}

In this section, we 
briefly review the description of stochastic chemical reaction systems using continuous-time Markov chains. 
We also review a reformulation of the chemical master equation using 
the language of quantum mechanics~\cite{Doi_1976, Doi_1976_2}.

\subsection{ Chemical reaction systems }

A chemical reaction network consists of 
the triple $(V,K,E)$
where 
$V$ is a set of chemical species, 
$K$ is a set of complexes, 
and 
$E$ is a set of chemical reactions. 
A complex is an element of $\mathbb N^V$,
where $\mathbb N$ denotes nonnegative integers,
and a reaction 
$e_A \in E$ is given by specifying 
two complexes as its source and target, 
\begin{equation}
e_A : 
  \sum_i  s_{iA} v_i 
  \longrightarrow 
  \sum_i t_{iA} v_i, 
\end{equation}
where $v_i \in V$. 
Here, $s_A, t_A \in \mathbb N^V$ are the source and target complexes of reaction $e_A$. 
A chemical reaction network can be represented as a directed graph of complexes, which is called a reaction graph. 
The reaction vector for $e_A$ is defined by 
$S_A \coloneqq t_A - s_A \in \mathbb Z^V$. 
Seen as a matrix, $S_{iA}$ is called a stoichiometric matrix.

For a given chemical reaction network, 
one can consider stochastic/deterministic dynamics on it. 
In the stochastic description, 
the variables that we use are the numbers of particles of chemical species, $n \in \mathbb N^V$. 
The status of a reaction system at time $t$ is 
characterized by the probability distribution of $n$, $P(t, n)$, and its time evolution is governed by the chemical master equation of a continuous-time Markov chain, 
\begin{equation}
\frac{d}{dt} 
p (t, n)
= 
\sum_A 
\left[ 
R_A (n - S_A) P(t, n-S_A) - R_A (n) P(t, n) 
\right] ,
\label{eq:cme}
\end{equation}
where $R_A (n)$ is the intensity function for reaction $e_A$.
Throughout the paper, 
we employ the mass-action kinetics,
\begin{equation}
R_A (n) = k_A \frac{n!}{(n-s_A)!},
\label{eq:mak}
\end{equation}
where $k_A$ is a constant. 
Here, the factorial of a vector $n \in \mathbb N^V$ is the abbreviation of the following expression, 
\begin{equation}
n! = \prod_i n_i !  . 
\end{equation}
The mass-action rate \eqref{eq:mak} is proportional to the number of combinations to form the source complexes for the reaction. 
This form is justified when the molecules in the system are well-stirred.

When the number of molecules is large and random fluctuations can be ignored, the system can be described by deterministic equations. 
In this case, the dynamical variables are the concentrations of chemical species, 
$x_i = n_i / V$, where $V$ is a parameter controlling the system size. 
The time evolution of $x_i$ is dictated by rate equations,
\begin{equation}
\frac{d}{dt} x_i (t) 
= 
\sum_A S_{iA} r_A (x),
\end{equation}
where $r_A (x)$ is the reaction rate of $e_A$. 
In the deterministic version of mass-action kinetics, reaction rates are written as 
\begin{equation}
r_A (x) = {\bar k}_A x^{s_A},
\label{eq:mak-det}
\end{equation}
where we have used the abbreviation $x^{s_A} = \prod_i x_i^{s_{iA}}$. 

In many situations, the stochastic description reduces to the deterministic one in the limit of a large system size.  
It is customary to take $R_A = O(V)$ and $n_i = O(V)$, where $V$ is a dimensionless parameter quantifying the system size. 
In this case, the parameters in the continuous-time Markov chain are related to those of the rate equation as 
\begin{equation}
\bar k_A = \lim_{V \to \infty } k_A V^{|s_A|_1 - 1} ,
\end{equation}
where $|s_A|_1$ denotes the L1 norm of $s_A$.

\subsection{ Quantum-mechanical formulation }

Quantum mechanics and stochastic chemical reaction systems have in common that the outcome of the measurement is probabilistic. 
With the quantum-mechanical formalism 
chemical reaction systems 
introduced in Refs. \cite{Doi_1976,Doi_1976_2}, 
the chemical master equation~\eqref{eq:cme}
is formally written as 
in the form of the the Schr\"{o}dinger equation. 
This gives us the opportunity to import the techniques established in the latter to the former problem, which is the strategy we take in this paper.
As we will see, this allowed us to find analytical form of  stationary distribution functions for chemical reaction networks that has not been known before to our knowledge. 

Below, we briefly review the quantum-mechanical formalism that we base on throughout this paper. (For a recent review of the quantum-mechanical formulation and the associated path-integral method~\cite{Peliti1985}, 
see Ref.~\cite{Weber_2017}.)
Let us start by introducing annihilation/creation operators for each species $v_i \in V$, 
$a_i$ and $a_i^\dag$, which obey the following commutation relations 
\begin{equation}
[a_i, a^\dag_j] 
= \delta_{ij}, 
\qquad
[a_i, a_j] =0, 
\qquad 
[a^\dag_i, a^\dag_j] =0, 
\label{eq:ccr}
\end{equation}
for any $i,j$. 
Roughly speaking, the creation (annihilation) operator $a_i^\dag$ ($a_i$) ``creates (annihilates)'' one species-$i$ molecule, as it would be clear in a moment.
We introduce the vacuum state, $\ket{0}$, 
as a state satisfying $a_i \ket{0} = 0$ and $\bra{0} a^\dag_i = 0$ for any $i$. 
Occupation number states are defined as
\begin{equation}
  \ket{n} := (a^\dag)^n \ket{0}, 
\end{equation}
where $n \in \mathbb N^V$ 
and we use the short-hand notation,
\begin{equation}
 (a^\dag)^{n} = \prod_i (a_i^\dag)^{n_i}. 
\end{equation}
As the name suggests, the vacuum state $\ket{0}$ and the occupation number state $\ket{n}$ describes a state where no molecules are present
%
and  a state that has occupation $n=(n_1,...,n_{|V|})$, respectively.
Note that the states are normalized as\footnote{
The Kronecker delta for 
$n,m \in \mathbb N^V$ is 
defined by $\delta_{n,m} = \prod_i \delta_{n_i,m_i}$. 
}
\begin{equation}
\langle n | m \rangle = n! \delta_{n,m},
\end{equation}
which is different from the one employed in quantum mechanics.

The actions of $a$ and $a^\dag$ on occupation number states are given by
\begin{equation}
  a^s \ket{n} = 
  \frac{n!}{(n-s)!}
  \ket{n-s}, 
  \qquad 
  (a^\dag)^s \ket{n} = \ket{n+s} ,
\end{equation}
for $s \in \mathbb N^V$. 
This relation makes it clear why $a$ and $a^\dag$ are called the annihilation and creation operators, respectively; when an annihilation (creation) operator $a_i$ ($a_i^\dag$) is applied to an occupation number state $\ket n$ for $s_i$ times, 
the number of species $i$ is decreased (increased) by $s_i$.

Using the occupation number states $\ket {n}$ defined above, we represent the probability distribution of the chemical reaction network 
at time $t$ as a vector $\ket{\psi(t)}$ as
\begin{equation}
  \ket{\psi (t)} = \sum_n P(t, n) \ket{n} . 
\end{equation}
Introducing the Hamiltonian $H$ by 
\begin{equation}
  H = 
  \sum_A 
  k_A 
  \left[ (a^\dag)^{t_A} -  (a^\dag)^{s_A} \right] 
  a^{s_A} ,
\label{eq:chem-hamiltonian}  
\end{equation}
the chemical master equation \eqref{eq:cme} 
can be expressed in the form of the Schr\"{o}dinger equation, 
\begin{equation}
  \frac{d}{dt} \ket{\psi(t)} = H \ket{\psi(t)} .
  \label{eq:s-eq}
\end{equation}
Indeed, one can check 
the equivalence of Eq.~\eqref{eq:s-eq} 
and Eq.~\eqref{eq:cme} by direct computation. 
In evaluating observables, the following state plays a special role, 
\begin{equation}
  \ket{\mathcal P} 
  := 
  e^{a^\dag} \ket{0}.
\end{equation}
This state satisfies 
$\bra{n} \mathcal P \rangle = 1$
\footnote{
  \begin{equation}
    \bra{n} \mathcal P \rangle
    = 
    \bra{0} a^n e^{a^\dag} \ket{0}
    = 
    \bra{0}e^{a^\dag} (a+1)^n \ket{0}
    = \langle 0 | 0 \rangle
    = 1. 
  \end{equation}
}
for any $\bra{n}$. 
For a given state $\ket{\psi}$, 
the expectation value of 
an observable $\mathcal O(n)$, 
which is a function of 
the numbers of molecules, is given by 
\begin{equation}
  \langle \mathcal O \rangle 
  = 
  \bra{\mathcal P} \mathcal O( a^\dag a )\ket{\psi(t)} , 
\end{equation}
so that the state $\ket{\psi(t)}$ represents a probability distribution, it should satisfy 
\begin{equation}
\bra{\mathcal P}\psi(t)\rangle = 1  ,
\label{eq:prob-cons}
\end{equation}
at any time $t$. 
If Eq.~\eqref{eq:prob-cons} is satisfied in the initial condition, it is also satisfied at later times, since 
\begin{equation}
\begin{split}    
  \bra{\mathcal P} H 
  &= 
  \sum_A 
  k_A
  \bra{0} 
  e^{a} 
  \left[ 
  (a^\dag)^{t_A} 
  - 
  (a^\dag)^{s_A} 
  \right] 
  a^{s_A}  \\
  &= 
  \sum_A 
  k_A
  \bra{0} 
  \left[ 
  (a^\dag+1)^{t_A} 
  - 
  (a^\dag+1)^{s_A} 
  \right] 
  e^{a} 
  a^{s_A}  \\
  &= 0 , 
\end{split}
\end{equation}
where we used the ``Doi shift,''
\begin{equation}
  e^a f(a^\dag) 
  = 
 f(a^\dag + 1) e^a, 
\end{equation}
and $\bra{0} a^\dag = 0$. 
We can see that 
the time evolution is consistent with 
probability conservation 
if the Hamiltonian $H = H(a, a^\dag)$ 
satisfies $H(a, a^\dag =1) = 0$.

\subsection{ Probability generating functions }

The formulation using creation/annihilation operators 
is equivalent to considering the time evolution of probability generating functions. 
The probability generating function is defined by 
\begin{equation}
\Psi (t,z) \coloneqq \sum_n P(t, n) z^n, 
\end{equation}
where $n \in \mathbb N^V$ and 
$z^n = \prod_i (z_i)^{n_i}$. 
To see the relation of the two formulations, 
note that $\frac{\p}{\p z_i}$ 
and $z_i$ satisfy the same 
commutation relations as Eq.~\eqref{eq:ccr}. 
The correspondence of the quantum-mechanical notation and the formulation based on generating functions 
can be made by the following replacements: 
\begin{equation}
\frac{\p}{\p z_i} \leftrightarrow a_i, 
\qquad 
z_i \leftrightarrow a_i^\dag, 
\qquad 
z^n  
\leftrightarrow 
\ket{n} . 
\end{equation}
Using probability generating functions, 
the chemical master equation can be written as~\cite{gardiner1985handbook}
\begin{equation}
\frac{d}{dt}
\Psi (t,z )
= 
\sum_A
k_A \left[ z^{t_A} - z^{s_A} \right] (\p_z)^{s_A} \Psi (t,z ). 
\label{eq:cme-gf}
\end{equation}
Note that we are using the following notations,
\begin{equation}
z^{t_A}
= 
\prod_i (z_i)^{t_{iA}}, 
\qquad 
(\p_z)^{s_A} = 
\prod_i 
\left( \frac{\p}{\p z_i} \right)^{s_{iA}} . 
\end{equation}
When all source complexes involve up to one species, the resulting equation for the probability generating function is a linear partial differential equation, that can be solved via the method of characteristics~\cite{doi:10.1073/pnas.0803850105}. 
An efficient method to obtain the analytic solution for these cases has recently been proposed~\cite{PhysRevE.104.024408}.

\subsection{ Anderson--Craciun--Kurtz theorem }

Finding the analytic form of stationary distributions is not easy in general. 
Anderson, Craciun, and Kurtz 
\cite{anderson2010product} showed that chemical master equations 
admit stationary distributions of a product-Poisson form when the deterministic counterpart (i.e. the rate equation) with the mass-action kinetics 
has a complex-balanced steady-state solution. 
In the quantum mechanical formulation of chemical master equations, these stationary distributions 
correspond to coherent states~\cite{baez2015quantum,baez2018quantum}.

A steady-state solution of the rate equation 
is said to be complex balanced 
when the following condition is satisfied 
\begin{equation}
\sum_{A: s_A =C_m} k_A \bar x^{s_A} 
= 
\sum_{A: t_A =C_m} k_A \bar x^{s_A} , 
\label{eq:cb-1}
\end{equation}
for any complex $C_m \in K$
in the reaction network. 
Intuitively, this means that the 
inflow and outflow of the rates are balanced in 
each complex. 
Equation~\eqref{eq:cb-1} 
can be written equivalently as 
\begin{equation}
\sum_A B_{mA} k_A {\bar x}^{s_A} =0,
\label{eq:cb-2}
\end{equation}
where $B_{mA}$ is the incidence matrix of 
the reaction graph. 
Note that the incidence matrix can be written 
using the Kronecker delta as 
$B_{mA} = \delta_{ t_A, C_m}  -  \delta_{ s_A, C_m}$\footnote{
The Kronecker delta of two complexes 
$s,s' \in \mathbb N^V$ should be understood as 
$\delta_{s, s'} \coloneqq \prod_i \delta_{s_{i}, s'_{i}}
$. 
},
and we have 
\begin{equation}
  \sum_A B_{m A} k_A \bar x^{s_A}
  = 
  \sum_{A} 
  ( \delta_{ t_A, C_m}  -  \delta_{ s_A, C_m} ) k_A \bar x^{s_A}
  = 
  \sum_{A : t_A = C_m} k_A \bar x^{s_A}
  - 
  \sum_{A : s_A = C_m} k_A \bar x^{s_A} . 
\end{equation}

Not every solution of rate equations 
has this property. 
An important class of reaction networks with complex-balanced steady states 
are those with a zero deficiency 
and weak reversibility. 
The deficiency $\delta$ is a nonnegative integer 
determined from the topological structure\footnote{
For an alternative approach to constrain the steady-state properties of deterministic chemical reaction systems with generic kinetics base on a different topological index, see Refs.~\cite{PhysRevLett.117.048101,PhysRevE.96.022322,PhysRevResearch.3.043123}. 
} of reaction networks, 
\begin{equation}
\delta \coloneqq |K| - \ell - {\rm rank}\, S, 
\end{equation}
where 
$|K|$ indicates the number of complexes, 
$\ell$ is the number of linkage classes (connected components of reaction graph),
and the last term is the rank of the stoichiometric matrix. 
A reaction network is said to be weakly reversible, 
if there is a path of reactions 
from one complex $C_m$ 
to another complex $C_n$,
there is always a path from $C_n$ to $C_m$. 
Feinberg~\cite{FEINBERG19872229} 
and Horn--Jackson \cite{horn1972general} 
showed that, if a reaction network has zero deficiency and is weakly reversible, 
the rate equation with mass-action kinetics 
admits a unique steady-state solution in each positive stoichiometric compatibility class for any choice of rate constants.

With a complex-balanced solution $\bar x \in \mathbb R^V$
of the rate equation with mass-action kinetics\footnote{
A tricky point here is that the rate equation is parametrized by $k_A$, which are the parameters of 
the stochastic reaction systems, and not those of the rate equation obtained by the deterministic limit \eqref{eq:mak-det} of the stochastic reaction system under consideration. 
In fact, the theorem holds even for 
non-mass-action stochastic kinetics \cite{anderson2010product,hong2021derivation},
whose deterministic limit does not have 
reaction rates with mass-action kinetics.
}, the Anderson--Craciun--Kurtz theorem claims that the following state is a stationary distribution\footnote{
Note that this is the abbreviation of the following expression,
\begin{equation}
\prod_i  e^{- \bar x_i} 
e^{ \bar x_i a^\dag_i }\ket{0} . 
\end{equation}
},
\begin{equation}
\ket{\bar x}
:= 
e^{- \bar x} 
e^{ \bar x a^\dag }\ket{0} ,
\label{eq:coh-st}
\end{equation}
where we have normalized the state 
so that $\bra{\mathcal P} \bar x \rangle =1$. 
This type of state is called a coherent state \cite{Gardiner_Zoller},
and it is an eigenstate of annihilation operators, 
\begin{equation}
  a_i \ket{\bar x} = \bar x_i \ket{\bar x} . 
\end{equation}

Let us explicitly show that 
the state~\eqref{eq:coh-st} 
is indeed a zero eigenstate of 
the chemical Hamiltonian~\eqref{eq:chem-hamiltonian}\footnote{
The following derivation is a slightly simplified version of the one given in Ref.~\cite{baez2015quantum}. 
See also Ref.~\cite{PhysRevE.96.062102}. 
}. 
To show this, a crucial step is 
writing the summation over reactions as 
\begin{equation}
  \sum_A
  = 
  \sum_m 
  \sum_{A: t_A =C_m }, 
\end{equation}
where $\sum_m$ is the summation over complexes, 
and $\sum_{A: t_A =C_m }$
is a summation over reactions whose reactants are given by complex $C_m$. 
Acting $H$ on the state \eqref{eq:coh-st}, 
\begin{equation}
\begin{split}
H 
e^{- \bar x} 
e^{\bar x a^\dag} \ket{0}
&=
e^{- \bar x}     
\sum_A 
k_A 
{\bar x}^{s_A} 
\left[  (a^\dag)^{t_A} -  (a^\dag)^{s_A} \right] 
    e^{\bar x a^\dag} \ket{0}
    \\  
&= 
e^{- \bar x}     
\sum_{m} 
\left[  \sum_{A: t_A =C_m }
 k_A  {\bar x}^{s_A}  (a^\dag)^{t_A} 
 - \sum_{A: s_A =C_m }
 k_A {\bar x}^{s_A} (a^\dag)^{s_A} 
 \right] 
    e^{\bar x a^\dag} \ket{0}
\\ 
&= 
e^{- \bar x}     
\sum_m 
(a^\dag)^{C_m} 
\left[ \sum_{A: t_A =C_m }
    k_A  {\bar x}^{s_A} 
    -  \sum_{A: s_A =C_m }
    k_A  {\bar x}^{s_A}  \right] 
    e^{\bar x a^\dag} \ket{0}
\\     
&= 
e^{- \bar x}     
\sum_m 
 (a^\dag)^{C_m} 
 \left( \sum_{A}  B_{mA} k_A  {\bar x}^{s_A}  \right) 
    e^{\bar x a^\dag} \ket{0}
 \\     
&=0, 
\end{split}
\end{equation}
where we used the complex-balancing condition, Eq.~\eqref{eq:cb-1} or \eqref{eq:cb-2}.

\section{ Squeezing and stochastic chemical reaction systems }\label{sec:squeeze-single}

We have seen that 
the stationary distributions of a product-Poisson form in the Anderson--Craciun--Kurtz theorem can be interpreted as 
coherent states in the quantum-mechanical formulation of stochastic chemical reaction systems. 
We consider the transformation of 
the Hamiltonian and the coherent state by a squeeze operator \cite{Gardiner_Zoller}. 
The transformed Hamiltonian represents a reaction network that is different from the original one. 
In particular, the transformed network has nonzero deficiency and is not weakly reversible. 
The obtained squeezed coherent state is 
the zero eigenstate (i.e. the stationary distribution) of the transformed Hamiltonian, and its analytic form can be identified. 
Therefore, although the transformed system is not of a zero deficiency and not weakly reversible in general, 
we can obtain the analytical expression 
for the stationary distribution through this procedure. 
In this section, we illustrate the procedure 
with a simple example.

\subsection{ Network transformation via squeezing }

We start with the following simple chemical reaction network, which we call $\Gamma$,
\begin{equation}
\begin{tikzcd}
\emptyset 
\rar["\ell_1", shift left] 
&
A
\lar["\ell_2", shift left]
\end{tikzcd} ,
\label{eq:network-0-a}
\end{equation}
where $\ell_1$ and $\ell_2$ are parameters 
in the mass-action kinetics of the corresponding reactions. 
The stochastic Hamiltonian for $\Gamma$ is given by
\begin{equation}
    H = 
    \ell_1  (a^\dag  - 1 ) 
    + 
    \ell_2  (1 - a^\dag) a 
    =     
    (a^\dag - 1) (\ell_1 - \ell_2 a)
    \eqqcolon 
    (a^\dag - 1)h(a), 
\end{equation}
where 
$a$ and $a^\dag$ are the creation 
and annihilation operators of species $A$, 
respectively, 
and we defined $h(a) \coloneqq \ell_1 - \ell_2 a$. 
The rate equation for this system is given by
\begin{eqnarray}
    \frac{dx}{dt} = \ell_1  - \ell_2 x, 
\end{eqnarray}
where $x(t)$ is the concentration of species $A$.

This network \eqref{eq:network-0-a} has zero deficiency and is weakly reversible. 
Hence, the Anderson--Craciun--Kurtz theorem applies, 
and a coherent state gives its stationary distribution. 
The steady-state solution of the rate equation 
is 
\begin{equation}
\bar x = \frac{\ell_1}{\ell_2}, 
\end{equation}
and, indeed, the state 
$\ket{\bar x} \coloneqq e^{- \bar x} e^{\bar x a^\dag}\ket{0}$ is a zero eigenstate of $H$, because 
\begin{eqnarray}
    h(a) \ket{\bar x}
    =(\ell_1 - \ell_2 a) \ket{\bar x}
    = (\ell_1 - \ell_2 \bar x) \ket{\bar x}
    = 0. 
\end{eqnarray}

We shall perform a squeezing and obtain 
another reaction network whose 
stationary distribution is given by a squeezed coherent state. A squeeze operator for $a$ is defined by \cite{Gardiner_Zoller}
\begin{equation}
S(\xi) = 
\exp 
\left[ 
\frac{1}2 \left( \xi^\ast a^2 - \xi (a^\dag)^2 \right) 
\right] ,
\end{equation}
where $\xi$ is a complex parameter. 
Under the action of $S(\xi)$, 
the operators $a$ and $a^\dag$ are transformed as\footnote{
The operator $S(\xi)$ is a unitary operator 
and $S^{-1}(\xi) = S^\dag(\xi) =S(-\xi)$. 
}
\begin{align}
S (\xi ) a S^{-1} (\xi)
&= 
\cosh r \, a + e^{i\theta } \sinh r \,a^\dag, 
\label{eq:s-a-s}
\\
S (\xi) a^\dag S^{-1} (\xi)
&=
\cosh r \, a^\dag
+ 
e^{-i\theta } \sinh r \, a , 
\label{eq:s-a-s-2}
\end{align}
where $\xi \coloneqq r e^{i\theta}$. 
Using the squeeze operator, 
we define a new Hamiltonian by 
\begin{eqnarray}
H'=(a^\dag - 1)S(\xi) h(a) S^{-1} (\xi). 
\end{eqnarray}
Namely, we have transformed the part that involves the annihilation operator, 
$h(a)$, by $S(\xi)$. 
The transformed Hamiltonian $H'$ 
is probability-conserving, 
since $H'(a,a^\dag = 1) = 0$. 
We can obtain a zero eigenstate of $H'$ by 
\begin{eqnarray}
\ket{\bar x,\xi} \coloneqq S(\xi)\ket{\bar x} . 
\label{eq:sq-state-def}
\end{eqnarray}
Namely, the zero eigenstate 
of the new Hamiltonian $H'$ is 
a squeezed coherent state. 
Indeed, we have 
\begin{eqnarray}
S(\xi) h(a) S^{-1}(\xi) \ket{\bar x,\xi}
= 
S(\xi) h(a) \ket{\bar x}
= 0. 
\end{eqnarray}
So that the state \eqref{eq:sq-state-def} 
represents a probability distribution, 
we take the parameter $\xi$ to be real. 
We will use the convention to take $\theta = 0$ and 
$r$ be of either sign. 
We will discuss more detailed properties of the stationary distribution given by Eq.~\eqref{eq:sq-state-def} in the next subsection.

Let us examine the chemical content of the transformed reaction system, which we call $\Gamma'$. 
Using Eq.~\eqref{eq:s-a-s}, $H'$ is written as 
\begin{equation}
H' 
= (a^\dag - 1)
[\ell_1 - \ell_2 (\cosh r \, a + \sinh r \, a^\dag)]. 
\end{equation}
The Hamiltonian can be organized in the following form, 
\begin{equation}
\begin{split}
H' 
&=
(\ell_1 + \ell_2 \sinh r) 
(a^\dag - 1) 
+ \ell_2 \cosh r (1- a^\dag) a 
- \ell_2 \sinh r ((a^\dag)^2 - 1) 
\\
&
\eqqcolon 
k_1 
(a^\dag - 1) 
+ 
k_2 (1- a^\dag) a 
+ 
k_3 ((a^\dag)^2 - 1) , 
\end{split}
\label{eq:hp-k}
\end{equation}
where we have defined 
\begin{equation}    
k_1 =  \ell_1 +\ell_2 \sinh r , 
\qquad 
k_2 = \ell_2 \cosh r ,
\qquad 
k_3 = - \ell_2 \sinh r . 
\end{equation}
Comparing Eq.~\eqref{eq:hp-k}
with a generic chemical Hamiltonian \eqref{eq:chem-hamiltonian},
we can see that the transformed Hamiltonian $H'$ corresponds to the following reaction network: 
\begin{equation}
\begin{tikzcd}
\emptyset 
\rar["k_1", shift left] 
\dar["k_3"] 
&
A 
\lar["k_2", shift left]
\\
2 A
& 
\end{tikzcd} 
\end{equation}
Compared to the original system $\Gamma$, a reaction $\emptyset \to 2A$ is added. 
The deficiency of the transformed network $\Gamma'$ is nonzero, $\delta=3-1-1=1$, and it is not weakly reversible. 

%
The parameters of $H$ 
can be expressed by those of $H'$ as 
\begin{equation}    
\ell_1 = k_1 + k_3, 
\qquad 
\ell_2 = \sqrt{(k_2)^2 - (k_3)^2} . 
\end{equation}
One might wonder if there are restrictions in the choice of the parameters $\{k_1, k_2, k_3\}$ 
from the positivity of $\{\ell_1, \ell_2\}$. 
In fact, there is no restriction and 
$\{k_1, k_2, k_3\}$ can be taken to be arbitrary positive values (some of them can be even zero). 
This is because the obtained probability generating function, once written in terms of the parameters 
$\{k_1, k_2, k_3\}$, is the solution of the stationary condition~\eqref{eq:cme-gf} 
of $\Gamma'$ for {\it any} positive values of $\{k_1, k_2, k_3\}$, even if some of $\ell_1,\ell_2$ are imaginary. 
In this sense, the reaction network $\Gamma$ is fictitious and is used as a stepping stone to compute the stationary distribution of $\Gamma'$. 
The fact that the stationary distribution of $\Gamma'$ from squeezing is indeed a stationary distribution can be 
checked independently of the properties of the original system $\Gamma$.

To simplify the notations, 
let us define 
\begin{equation}
u \coloneqq \frac{k_1}{k_2},
\qquad 
v \coloneqq \frac{k_3}{k_2},
\qquad 
\gamma \coloneqq \frac{1}{\sqrt{1-v^2}}. 
\end{equation}
The squeezing parameter is given by 
$\tanh r = -v$, 
and $\bar x$ is written as $\bar x = \gamma (u + v)$. 


\subsection{ Stationary distribution }

We can utilize the mapping of $\Gamma$ 
and $\Gamma'$ to compute the exact stationary distribution of $\Gamma'$, 
which has a nonzero deficiency and 
is not weakly reversible: 
the stationary distribution of $\Gamma'$ is given by the squeezed coherent state~\eqref{eq:sq-state-def}. 
To find the probability distribution, 
let us here use the representation using probability generating functions. 
The reaction system $\Gamma$ has 
a Poisson distribution as its stationary distribution, 
and the corresponding probability generating function is written as 
\begin{equation}
\Psi_{\rm c}(z) 
= 
e^{-\bar x} 
\sum_n \frac{{\bar x}^n}{n!}  z^n 
= 
e^{\bar x (z-1)}. 
\end{equation}
It is an eigenfunction of $\p_z$
with eigenvalue $\bar x$, 
\begin{equation}
\p_z 
\Psi_{\rm c} (z)
= 
\bar x \Psi_{\rm c} (z). 
\label{eq:mom-gen-c-eigen}
\end{equation}
The probability generating function of 
the stationary distribution of 
$\Gamma'$ 
is obtained by operating a squeeze operator
on Eq.~\eqref{eq:mom-gen-c-eigen}, 
\begin{equation}
\Psi_{\rm sq}(z) 
= 
S(\xi)\Psi_{\rm c}(z) 
=
e^{\frac{1}2 (\xi^\ast (\p_z)^2 - \xi z^2)} 
e^{\bar x (z-1)}. 
\label{eq:mom-gen-sq}
\end{equation}

Although it is possible to compute 
Eq.~\eqref{eq:mom-gen-sq} directly, 
let us take an easier path. 
Here, we use the eigenvalue equation satisfied by $\Psi_{\rm sq}(z)$. 
Acting $S(\xi)$ on both sides
of Eq.~\eqref{eq:mom-gen-c-eigen}, 
\begin{equation}
S(\xi) \p_z S^{-1}(\xi) 
\Psi_{\rm sq} (z)
= 
\bar x
\Psi_{\rm sq} (z). 
\end{equation}
Namely, 
$\Psi_{\rm sq} (z)$ is 
the eigenfunction of the operator 
$
S(\xi) \p_z S^{-1}(\xi) 
$
with eigenvalue $\bar x$. 
The operator $S(\xi) \p_z S^{-1}(\xi)$ is written as 
\begin{equation}
S(\xi)
\p_z 
S^{-1}(\xi) 
= 
\cosh r \, \p_z + \sinh r  \, z ,
\end{equation}
which is equivalent to Eq.~\eqref{eq:s-a-s}. 
Thus, we have a differential equation,
\begin{equation}
( \p_z + \tanh r \, z ) 
\Psi_{\rm sq} (z)
= 
\frac{\bar x}{\cosh r}  
\Psi_{\rm sq} (z). 
\end{equation}
The solution can be readily obtained as 
\begin{equation}
\Psi_{\rm sq}(z) 
= 
\exp 
\left[
- \frac{\tanh r}{2} (z^2 -1) 
+\frac{\bar x }{\cosh r} (z-1)
\right] 
= 
\exp 
\left[
\frac{v}{2} (z^2 -1) + (u+v) (z-1)
\right] .
\label{eq:psi-sq-analytic}
\end{equation}
where we used the parameters of $\Gamma'$ 
in the last expression. 
We have fixed the normalization constant by the condition 
$
\Psi_{\rm sq}(z =1) = 1
$.

The probability generating function \eqref{eq:psi-sq-analytic} fully characterizes the stationary distribution, 
and we can use this to evaluate the statistical properties of $\Gamma'$ in the long-time limit. 
For example, we can compute cumulants using 
the cumulant generating function, 
\begin{equation}
C(w) \coloneqq \ln \Psi_{\rm sq}(e^w)
= 
\frac{v}{2} (e^{2w} -1) 
+
(u+v) (e^{w}-1). 
\end{equation}
The $n$-th cumulant, $c_n$, is computed as 
\begin{equation}
c_n = 
u + 
( 2^{n-1} + 1 ) v
= 
\frac{k_1}{k_2} 
+ 
( 2^{n-1} + 1 ) \frac{k_3}{k_2} . 
\end{equation}
For example, the mean and variance are 
\begin{align}
&\langle n \rangle 
= 
u+2v
=
\frac{k_1 + 2 k_3}{k_2} ,
\\
&\langle n^2 \rangle - \langle n \rangle^2
= 
\langle n \rangle + v 
=
\frac{k_1 + 3 k_3}{k_2} . 
\end{align}
Recalling that a Poisson distribution has identical cumulant for arbitrary $n$,
one sees that the stationary distribution 
in the transformed system 
is broader than a Poisson distribution.

To get the expression of the stationary 
number distribution, note that the generating function of the Hermite polynomials is 
given by 
\begin{equation}
e^{- z^2 + 2 y z} 
= 
\sum_{n}
\frac{1}{n!}
H_n (y) 
z^n . 
\end{equation}
Using this, the probability generating function can be expanded as 
\begin{equation}
\begin{split}
\Psi_{\rm sq}(z) 
&= 
\exp 
\left(
\frac{\tanh r}{2} 
- \frac{\bar x}{\cosh r} 
\right) 
\sum_n 
\frac{1}{n!}
\left( 
\frac{ \tanh r }{2}
\right)^{\frac{n}2}
H_n
\left( \frac{\bar x }{\sqrt{\sinh 2 r}} \right) 
\, 
z^n 
\\
&= 
e^{ - \frac{3}{2} v - u }
\sum_n 
\frac{1}{n!}
\left( 
- \frac{v}{2} 
\right)^{\frac{n}2}
H_n
\left( \frac{u+v}{\sqrt{ -2 v}} \right) 
\, 
z^n . 
\end{split}
\end{equation}
This expression coincides with 
the photon number distribution 
of squeezed coherent states~\cite{Gong1990} 
up to normalization. 
From the coefficients, 
we can read off the stationary distribution $P_{\rm s} (n) $, which can be expressed
using $\{k_1,k_2,k_3\}$ as 
\begin{equation}
P_{\rm s} (n) 
= 
e^{- \frac{1}{k_2} 
\left(
\frac{3}{2}k_3+ k_1  \right) 
}
\frac{1}{n!}
\left( 
- \frac{k_3}{2k_2 } 
\right)^{\frac{n}2}
H_n
 \left( \frac{k_1+k_3}{\sqrt{ -2 k_2 k_3}} \right) .
 %
\label{eq:eg1-prob-analytic}
\end{equation}

\begin{figure}[tb] 
\centering
\includegraphics[width=8cm]{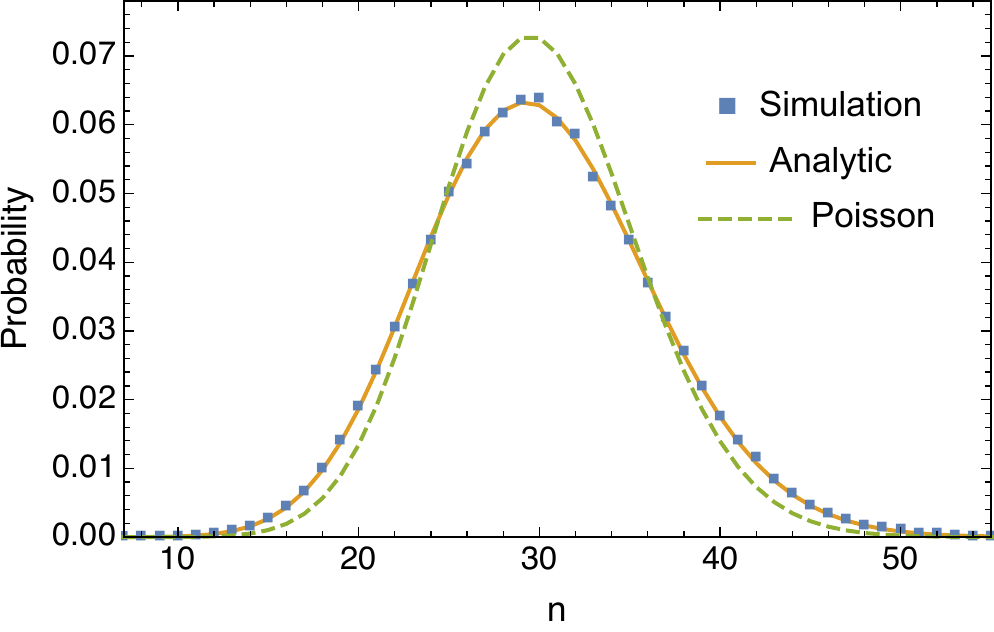} 
\caption{ 
Probability distributions from 
the analytic solution and numerical simulations. 
The solid line shows the analytic solution, 
which is consistent with numerically simulations (boxes). 
The distribution is wider than the Poisson distribution (dashed line) with the same mean. 
Parameters are chosen as $k_1=10, k_2=1, k_3=10$.  
The numerical data are based on $200,000$ simulated events. 
}
\label{fig:eg1-comp}
\end{figure}

In Fig.~\ref{fig:eg1-comp}, we show a comparison of the analytic form of the stationary distribution~\eqref{eq:eg1-prob-analytic} with stochastic simulations using the Gillespie algorithm~\cite{gillespie1977exact}. 
The numerically computed distribution agrees well with the analytic expression.

\section{ Example with correlations from two-mode squeezing }\label{sec:two-mode-sq}

A noticeable feature of stationary distributions for complex-balanced systems is that they are of a product form and each species is statistically independent when there are no conserved quantities. 
Here, we discuss an example where 
the transformed reaction system has 
a correlated stationary distribution among different species, which is obtained by the so-called two-mode squeezing used in the context of continuous variable-quantum information processing \cite{Weedbrook2012}.

\subsection{ Network transformation }

As an original network $\Gamma$, we consider the following: 
\begin{equation}
\begin{tikzcd}
A
\rar["\ell_2", shift left] 
&
\emptyset 
\lar["\ell_1", shift left]
\rar["\ell_3", shift left] 
& 
B
\lar["\ell_4", shift left] 
\end{tikzcd}
\end{equation}
%
%
The corresponding Hamiltonian reads 
\begin{equation}
H = 
\ell_1 (1 - a^\dag) a 
+ \ell_2 (a^\dag - 1) 
+\ell_3 (1 - b^\dag) b 
+ \ell_4 (b^\dag - 1) 
= 
 (a^\dag-1) 
 (\ell_1 - \ell_2 a) 
 + (b^\dag-1) 
 (\ell_3 - \ell_4 b),
\end{equation}
where $a(a^\dag)$ and $b(b^\dag)$ 
are the annihilation (creaction) 
operators for species $A$ and $B$, respectively. 
This network is weakly reversible and its deficiency is zero, 
so the rate equations admit a complex-balanced steady-state solution.
The rate equations are 
\begin{equation}    
\frac{d}{dt} x_a = \ell_1 - \ell_2 x_a,
\qquad 
\frac{d}{dt} x_b = \ell_3 - \ell_4 x_b,
\end{equation}
where $x_a(t)$ and $x_b(t)$ are concentrations of species $A$ and $B$, 
respectively. 
The steady-state concentrations are 
given by
\begin{equation}
\bar x_a = \frac{\ell_1}{\ell_2}, 
\qquad 
\bar x_b = \frac{\ell_3}{\ell_4}. 
\label{eq:xa-xb}
\end{equation}
The stationary state of this system is the product of Poisson distributions with parameters \eqref{eq:xa-xb}. 
On the stationary state 
and Hamiltonian of this system, 
we act a two-mode squeeze operator, 
\begin{equation}
S_2 (\zeta) = \exp \left( 
\zeta^\ast a b - \zeta a^\dag b^\dag \right) ,
\end{equation}
which mixes the operators of different species. 
The operators $a$ and $b$ 
are transformed as 
\begin{align}
a 
\mapsto 
S_2 (\zeta) a S^{-1}_2 (\zeta)
&= \cosh q \,  a + e^{i \phi} \sinh q \,  b^\dag, 
\\
b 
\mapsto 
S_2 (\zeta) b S^{-1}_2 (\zeta)
&= \cosh q \, b +  e^{i \phi} \sinh q\,  a^\dag ,
\end{align}
where $\zeta = q e^{i\phi}$.

We define the transformed Hamiltonian by 
\begin{equation}
H' = (a^\dag-1) 
S_2 (\zeta)(\ell_1 - \ell_2 a)S^{-1}_2 (\zeta)
+ 
(b^\dag-1) 
S_2 (\zeta)(\ell_3 - \ell_4 b) S^{-1}_2 (\zeta)
,
\end{equation}
where we take $\phi = 0$, and $q$ can be either positive or negative. 
To read off its chemical content, 
let us rewrite the Hamiltonian as 
\begin{equation}
\begin{split}
H' &=(a^\dag-1) 
 (\ell_1 - \ell_2 \cosh q\, a - \ell_2 \sinh q\, b^\dag) 
 + (b^\dag-1) 
 (\ell_3 - \ell_4 \cosh q\, b - \ell_4 \sinh q\, a^\dag) 
 \\
&=(a^\dag-1) 
 (\ell_1 - \ell_2 \cosh q\, a) 
 + (b^\dag-1) 
 (\ell_3 - \ell_4 \cosh q\, b)
 -\ell_2 \sinh q(a^\dag b^\dag 
 \textcolor{blue}{-1}- b^\dag 
 \textcolor{blue}{+1})  
 - \ell_4 \sinh q (a^\dag b^\dag \textcolor{blue}{-1}-a^\dag\textcolor{blue}{+1})
    \\
&=(a^\dag-1) 
 ((\ell_1+
 {\ell_4 \sinh q}) - \ell_2 \cosh q \, a)
+ (b^\dag-1) 
 ((\ell_3
 {+\ell_2 \sinh q}) - \ell_4 \cosh q\, b)
 -(\ell_2+\ell_4) \sinh q(a^\dag b^\dag-1) 
\\
&\eqqcolon 
(a^\dag-1) (k_1 - k_2 a)
+ (b^\dag-1) (k_3 - k_4 b)
+ k_5(a^\dag b^\dag-1),
\end{split}    
\end{equation}
where we have inserted $+1 -1$ (colored in blue) in the second line. 
Comparing this with the form of a generic Hamiltonian \eqref{eq:chem-hamiltonian}, 
the transformed reaction system corresponds to a network $\Gamma'$ with the following reactions,
\begin{equation}
\begin{tikzcd}
A
\rar["k_2", shift left] 
&
\emptyset 
\lar["k_1", shift left]
\rar["k_3", shift left] 
\dar["k_5"] 
& 
B
\lar["k_4", shift left] 
\\
&
A+B
&
\end{tikzcd} 
\end{equation}
The deficiency of this network is one, $\delta = 1$, and is not weakly reversible. 
The parameters of $\Gamma'$ are written 
by those of $\Gamma$ as 
\begin{equation}    
k_1 = \ell_1+\ell_4 \sinh q  
\quad 
k_2 = \ell_2\cosh q, 
\quad 
k_3 = \ell_3+\ell_2 \sinh q, 
\quad 
k_4 = \ell_4\cosh q, 
\quad 
k_5 = -(\ell_2 + \ell_4)\sinh q. 
\end{equation}
Note that $k_1$ and $k_3$ 
can be taken to be zero, 
in which case the corresponding reaction 
is absent in the network. 
%
%
The parameters of $\Gamma$ can 
be expressed by 
the parameters of $\Gamma'$ as 
\begin{equation}
\ell_1 = k_1 + v k_4, 
\qquad 
\ell_2 = k_2 \sqrt{1-v^2}, 
\qquad 
\ell_3 = k_3 + v k_2, 
\qquad 
\ell_4 = k_4 \sqrt{1-v^2},    
\end{equation}
where we have defined 
\begin{equation}
v \coloneqq \frac{k_5}{k_2 + k_4}. 
\end{equation}
The parameter $q$ is also determined from 
the parameters $k_i$
as $\tanh q = -v$. 
%

A similar comment to the previous example also applies here. 
For some choice of the parameters 
$\{k_i\}_{i=1\ldots5}$,
some of $\{\ell_i \}_{i=1\ldots4}$ can become imaginary. 
However, the stationary distribution 
obtained through squeezing in fact is correct for {\it any} positive values 
$\{k_i\}_{i=1\ldots5}$, 
because the squeezed coherent state is 
going to be the zero mode of $H'$ 
regardless of whether $\ell_i$ are real or imaginary.

\subsection{ Stationary distribution }

Here, we look at the properties 
of the stationary distribution. 
For this purpose, 
we will use the probability generating function. 
The original state is a coherent state 
and the corresponding 
probability generating function is written as 
\begin{equation}
\Psi_{\rm c} (z_a, z_b)
= 
e^{\bar x_a (z_a - 1)}
e^{\bar x_b (z_b - 1)}. 
\label{eq:coh-s-corr}
\end{equation}
We will denote the derivatives 
with respect to $z_a$ and $z_b$ as 
\begin{equation}
\p_a \coloneqq \frac{\p}{\p z_a}, 
\qquad 
\p_b \coloneqq \frac{\p}{\p z_b}. 
\end{equation}
Equation~\eqref{eq:coh-s-corr} is an eigenfunction of derivative operators, 
\begin{equation}    
\p_a 
\Psi_{\rm c} (z_a, z_b)
= 
\bar x_a \Psi_{\rm c} (z_a, z_b), 
\qquad 
\p_b
\Psi_{\rm c} (z_a, z_b)
= 
\bar x_b \Psi_{\rm c} (z_a, z_b). 
\label{eq:coh}
\end{equation}
The zero eigenstate of the transformed reaction system is obtained by acting $S_2(\zeta)$ 
on $\Psi_{\rm c}(z_a,z_b)$, 
\begin{equation}
\Psi_{\rm sq}  (z_a, z_b)
\coloneqq 
S_2 (\zeta) \Psi_{\rm c}(z_a, z_b). 
\end{equation}
Similarly to the case of single-mode squeezing, 
to find the expression of $\Psi_{\rm sq}  (z_a, z_b)$, we use the following eigenvalue equations
obtained by acting $S_{2}(\zeta)$ 
on Eq.~\eqref{eq:coh}, 
\begin{align}
S_2 (q) \p_a S^{-1}_2 (q)
\Psi_{\rm sq} (z_a, z_b)
&= 
\bar x_a \Psi_{\rm sq} (z_a, z_b), 
\label{eq:sps-sq-sq-a}
\\
S_2 (q) \p_b S^{-1}_2 (q)
\Psi_{\rm sq} (z_a, z_b)
&= 
\bar x_b \Psi_{\rm sq} (z_a, z_b),
\label{eq:sps-sq-sq-b}
\end{align}
where we have taken the parameter to be real, 
$\zeta = q \in \mathbb R$. 
Namely, $\Psi_{\rm sq}(z_a,z_b)$ 
is an eigenfunction of 
operators, 
$S_2 (q) \p_a S^{-1}_2 (q)$
and 
$S_2 (q) \p_b S^{-1}_2 (q)$. 
Noting that 
\begin{equation}
S_2 (q) \p_a S^{-1}_2 (q)
= 
\cosh q \, \p_a + \sinh q \, z_b, 
\qquad 
S_2 (q) \p_b S^{-1}_2 (q)
= 
\cosh q \, \p_b + \sinh q \, z_a, 
\end{equation}
Eqs.~\eqref{eq:sps-sq-sq-a}
and \eqref{eq:sps-sq-sq-b} 
are written as 
\begin{align}
(\p_a + \tanh q \, z_b)\Psi_{\rm sq} (z_a, z_b)
&= 
\frac{\bar x_a}{\cosh q} \Psi_{\rm sq} (z_a, z_b), 
\\
(\p_b + \tanh q \, z_a)\Psi_{\rm sq} (z_a, z_b)
&= 
\frac{\bar x_b}{\cosh q} \Psi_{\rm sq} (z_a, z_b). 
\end{align}
These differential equations can be immediately solved to give 
\begin{equation}
\begin{split}
\Psi_{\rm sq} (z_a, z_b)
&= 
\exp 
\left( 
- \tanh q (z_a z_b -1 )
+ \frac{\bar x_a}{\cosh q} (z_a -1)
+ \frac{\bar x_b}{\cosh q} (z_b -1) 
\right) 
\\
&=
\exp 
\left( 
v (z_a z_b -1 )
+ 
\frac{k_1 + v k_4}{k_2} (z_a -1)
+ 
\frac{k_3 + v k_2}{k_4} (z_b -1) 
\right) ,
\end{split}
\end{equation}
where we used the parameters of $\Gamma'$ 
in the second line. 
We have fixed the normalization constant 
using the condition 
$\Psi_{\rm sq}(z_a=1,z_b=1)=1$.

The probability generating functions of 
marginalized distributions for $n_a$ and $n_b$ are 
given by 
\begin{align}
\psi_a (z_a)
\coloneqq  
\Psi_{\rm sq} (z_a, z_b = 1)
&= 
\exp 
\left[
\frac{k_1 + k_5}{k_2}
(z_a -1)
\right] ,
\\
\psi_b (z_b)
\coloneqq  
\Psi_{\rm sq} (z_a=1, z_b)
&= 
\exp 
\left[
\frac{k_3 + k_5}{k_4} (z_b -1)
\right] 
.  
\end{align}
Thus, marginal distributions are Poissonian, 
and they are characterized by 
the following parameters: 
\begin{equation}
\langle n_a \rangle 
= 
\frac{k_1 + k_5}{k_2},
\qquad 
\langle n_b \rangle 
= 
\frac{k_3 + k_5}{k_4}. 
\end{equation}
As a result of a two-mode squeezing, 
the joint distribution is not a product of Poisson distributions, 
and $n_a$ and $n_b$ are correlated. 
The covariance is given by
\begin{equation}
\langle n_a n_b \rangle 
- 
\langle n_a \rangle 
\langle n_b \rangle 
= 
v
= \frac{k_5}{k_2 + k_4} . 
\end{equation}

\subsection{ Derivation of the number distribution }

Here, we derive the analytic expression of 
the stationary distribution of this reaction system. 
Let us write the probability generating function in the following form
\begin{equation}
\Psi_{\rm sq} (z_a, z_b)
= 
\exp 
\left[ 
v  (z_a z_b -1 )
+ c_a (z_a -1) + c_b (z_b -1) 
\right] , 
\end{equation}
where we defined 
\begin{equation}
c_a \coloneqq \frac{k_1 + v k_4}{k_2} ,
\qquad 
c_b \coloneqq \frac{k_3 + v k_2}{k_4}    .
\end{equation}
Expanding the generating function in powers of $z_a$ and $z_b$, 
\begin{equation}
\begin{split}
\Psi_{\rm sq} (z_a, z_b)
&= 
e^{ - v - c_a - c_b} 
\sum_{k,l,m}
\frac{1}{k! \, l! \, m!} 
v^k (c_a)^l (c_b)^m 
(z_a)^{k+l} (z_b)^{k+m}
\\
&= 
e^{ - v - c_a - c_b}
\sum_{n_a, n_b} 
\sum_{k,l,m}
\delta_{k+l,n_a}
\delta_{k+m,n_b}
\frac{v^k (c_a)^l (c_b)^m }{k! \, l! \, m!} 
(z_a)^{k+l} (z_b)^{k+m}
\\
&= 
e^{ - v - c_a - c_b}
\sum_{n_a, n_b} 
\sum_{k=0}^{\min (n_a,n_b)} 
\frac{v^k (c_a)^{n_a - k} (c_b)^{n_b - k} }{k! \, (n_a-k)! \, (n_b - k)!} 
(z_a)^{n_a} (z_b)^{n_b} . 
\end{split} 
\end{equation}
We can read off the number distribution 
from the coefficient of $(z_a)^{n_a} (z_b)^{n_b}$. 
%
%
To further simplify the expression, 
let us first consider the case 
$n_a \le n_b$: 
\begin{equation}
\begin{split}
P_{\rm s}(n_a,n_b)
&= 
e^{ - v - c_a - c_b}
\sum_{k=0}^{n_a}
\frac{v^k (c_a)^{n_a - k} (c_b)^{n_b - k} }{k! \, (n_a-k)! \, (n_b - k)!} 
\\
&= 
\frac{e^{ - v - c_a - c_b}}{n_b !}
v^{n_a} (c_b)^{n_b - n_a}
\sum_{k=0}^{n_a}
\frac{n_b ! }{k! \, (n_a-k)! \, (n_b - n_a + k)!} 
\left(
\frac{c_a c_b}{v}
\right)^k 
\\
&= 
\frac{e^{ - v - c_a - c_b}}{n_b !}
 v^{n_a} (c_b)^{n_b - n_a}
L_{n_a}^{(n_b - n_a )} 
\left(- \frac{c_a c_b}{v} \right) ,
\end{split}
\end{equation}
where 
$L_{n}^{(p)}\left(x \right)$ 
are generalized Laguerre polynomials,
and we changed the summation label 
$k \leftrightarrow (n_a - k)$ in the second line. We also used the following expression of generalized Laguerre polynomials
\begin{equation}
L_n^{(p)} (x)
= \sum_{k=0}^{n} \frac{(n+p)!}{(p+k)!(n-k)!k!} (-y)^k. 
\end{equation}
\begin{figure}[tb] 
\centering
\includegraphics[width=8cm]{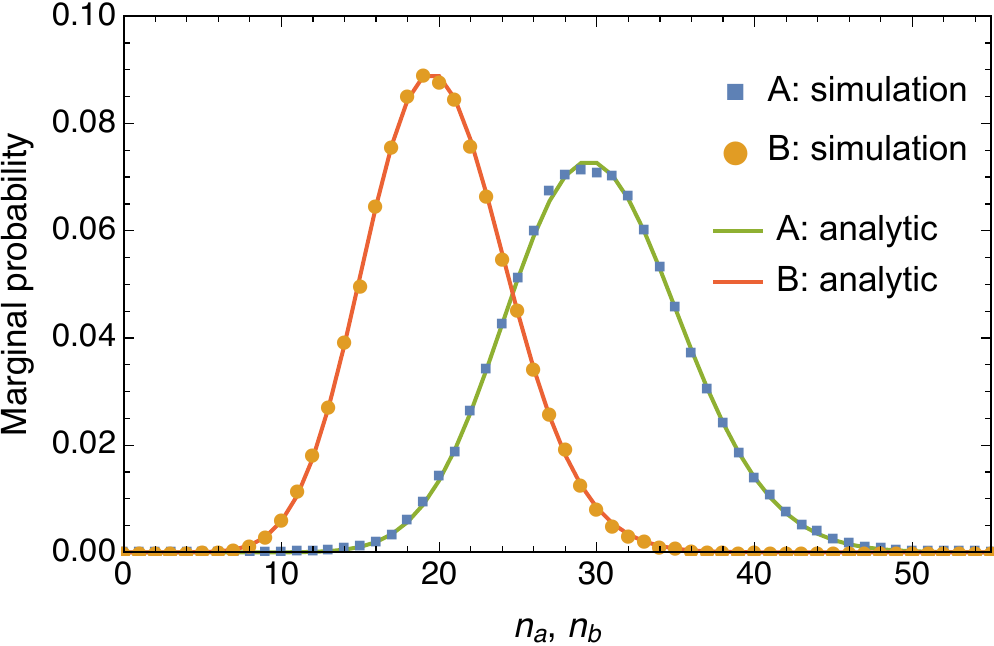} \caption{ 
Marginal probability distributions of $n_a$ and $n_b$ from the analytic solution and numerical simulations. 
Solid lines are from the analytic solution,
which agrees well with the result of numerical simulations indicated by boxes (species A) and circles (species B). 
The plot is based on $100,000$ events. 
Parameters are chosen as $k_1=20,k_2=1,k_3=10,k_4=1,k_5=10$. 
}
\label{fig:eg-corr-marginal}
\end{figure}
\begin{figure}[tb] 
\centering
\includegraphics[width=9cm]{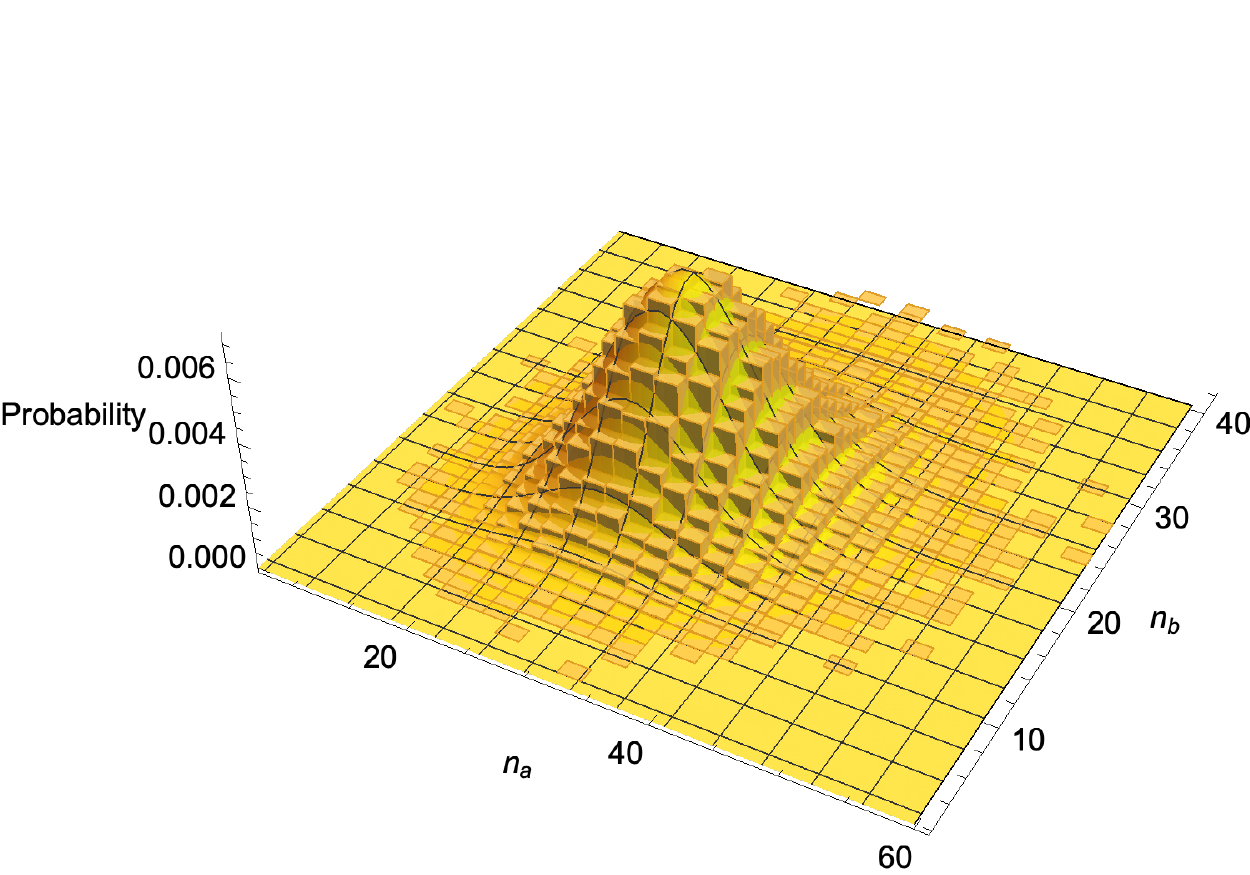} 
\caption{ 
Joint probability distributions from 
the analytic solution and numerical simulations. 
Parameters and the number of events are the same as Fig.~\ref{fig:eg-corr-marginal}.
}
\label{fig:eg-corr-joint}
\end{figure}
A similar expression can be 
obtained for the case $n_a \ge n_b$. 
Introducing 
$p \coloneqq \min (n_a,n_b)$
and 
$q \coloneqq \max (n_a,n_b)$, 
the stationary distribution is finally written as 
\begin{equation}
P_{\rm s}(n_a,n_b)
= 
\frac{ e^{-v-c_a-c_b} }{q !}
v^p (c_a)^{n_a - p} (c_b)^{n_b - p}
L_{p}^{(q-p)} 
\left( - \frac{c_a c_b}{v} \right) . 
\end{equation}
The expression matches with the coefficients 
of two-mode squeezed states \cite{PhysRevA.43.3854} (up to a normalization constant). 

We have validated the analytically computed probability distributions
with numerically computed ones 
using the Gillespie algorithm. 
In Fig.~\ref{fig:eg-corr-marginal}, 
we plot the analytically calculated marginal distributions of species $A$ and $B$, 
which are consistent with numerical simulations. 
Figure~\ref{fig:eg-corr-joint} shows the analytic form of the joint distribution 
and a histogram based on Monte Carlo simulations. The two are consistent.

\section{ Example with non-linear reactions }\label{sec:nonlinear}

The present method can be applied 
to reaction systems that contain 
reactions whose source complexes involve two or more species. 
In such cases, the differential equations satisfied by the generating functions (in the stationary state) contain two or more partial derivatives, and obtaining analytic solutions 
is a nontrivial task. 
Squeezing can be applied in such situations as well.
Here, we discuss such an example.

As a starting point, let consider the following network $\Gamma$,
\begin{equation}
\label{eq:network 0 B A+B}
\begin{tikzcd}
\emptyset
\rar["\ell_3", shift left] 
&
B
\lar["\ell_4", shift left]
\rar["\ell_1", shift left] 
&
A + B
\lar["\ell_2", shift left]
\end{tikzcd} 
\end{equation}
The reaction $A+B \to B$ involves two species. 
The network $\Gamma$ is of a zero deficiency and weakly reversible. 
Hence, the stationary distribution 
is a product of Poisson distributions,
whose means are given by the steady-state solution, 
\begin{equation}
\bar x_a = \frac{\ell_1}{\ell_2} , 
\qquad 
\bar x_b = \frac{\ell_3}{\ell_4} .
\end{equation}
The stochastic Hamiltonian is written as 
\begin{equation}
\begin{split}
H 
&= 
\ell_1 
( a^\dag b^\dag - b^\dag ) 
b
+ 
\ell_2 ( b^\dag - a^\dag b^\dag ) 
a b
+ \ell_3 ( b^\dag - 1) 
+ \ell_4 (1-b^\dag) b
\\
&= 
( a^\dag b^\dag - b^\dag ) 
(\ell_1 b - \ell_2 a b )
+ 
( b^\dag - 1) (\ell_3  - \ell_4 b) .
\end{split}
\end{equation}
On this system, we perform the single-mode squeezing of species $A$. 
The annihilation and creation operators of species $A$ 
are transformed as Eqs.~\eqref{eq:s-a-s} and \eqref{eq:s-a-s-2}, respectively. 
The transformed Hamiltonian reads 
\begin{equation}
\begin{split}
H' 
&= 
( a^\dag b^\dag - b^\dag ) 
S_a(r)
(\ell_1 b - \ell_2 a b )
S_a^{-1}(r) 
+ 
( b^\dag - 1) (\ell_3  - \ell_4 b) 
\\
&= 
( a^\dag b^\dag - b^\dag ) 
(\ell_1 b - \ell_2 \cosh r\, a b )
+ 
( b^\dag - 1) (\ell_3  - \ell_4 b) 
- 
\ell_2 \sinh r
( (a^\dag)^2 b^\dag - a^\dag b^\dag ) b . 
\end{split}
\end{equation}
We write the last term as 
\begin{equation}
- 
\ell_2 \sinh r 
( (a^\dag)^2 b^\dag - b^\dag - a^\dag b^\dag + b^\dag) b 
=     
- \ell_2 \sinh r
( (a^\dag)^2 b^\dag - b^\dag ) b
+ 
\ell_2 \sinh r ( a^\dag b^\dag - b^\dag) b .
\end{equation}
Hence, the total Hamiltonian is written as 
\begin{equation}
\begin{split}
H'
&=
( a^\dag b^\dag - b^\dag ) 
((\ell_1 + \ell_2 \sinh r) b - \ell_2 \cosh r\, a b )
+ 
( b^\dag - 1) (\ell_3  - \ell_4 b) 
- \ell_2 \sinh r
( (a^\dag)^2 b^\dag - b^\dag ) b
\\
&\eqqcolon
( a^\dag b^\dag - b^\dag ) 
(k_1 b - k_2 a b )
+ 
( b^\dag - 1) (k_3  - k_4 b) 
+ 
k_5 ( (a^\dag)^2 b^\dag - b^\dag ) b .
\end{split}
\end{equation}
This Hamiltonian represents the following reaction network, 
\begin{equation}
\label{eq:network 0 B A+B 2A+B}
\begin{tikzcd}
\emptyset
\rar["k_3", shift left] 
&
B
\lar["k_4", shift left]
\rar["k_1", shift left] 
\dar["k_5"] 
&
A+B
\lar["k_2", shift left]
\\
&
2 A + B
& 
\end{tikzcd} 
\end{equation}
which we call $\Gamma'$. 
The parameters of $\Gamma'$ is written 
using those of $\Gamma$ as 
\begin{equation}
k_1 = \ell_1 + \ell_2 \sinh r, 
\quad 
k_2 = \ell_2 \cosh r, 
\quad 
k_3 = \ell_3, 
\quad 
k_4 = \ell_4,
\quad 
k_5 = - \ell_2 \sinh r .
\end{equation}
We can solve this for $\{\ell_i\}_{i=1\ldots4}$ and $r$ as 
\begin{equation}
\ell_1 = k_1 + k_5, 
\quad 
\ell_2 = \sqrt{(k_2)^2 - (k_5)^2}, 
\quad 
\tanh r = - \frac{k_5}{k_2}. 
\end{equation}
The probability generating function $\Psi'(z_a,z_b)$ 
for the system $\Gamma'$ satisfies the following differential equations,
\begin{align}
\left( \p_a 
- \frac{k_5}{k_2} 
z_a \right) \Psi' (z_a,z_b) 
&= 
\frac{k_1 + k_5}{k_2}
\Psi' (z_a,z_b), 
\\
\p_b \Psi' (z_a,z_b) &= 
\frac{k_3}{k_4}
\Psi' (z_a,z_b) .
\end{align}
\begin{figure}[tb] 
\centering
\includegraphics[width=8cm]{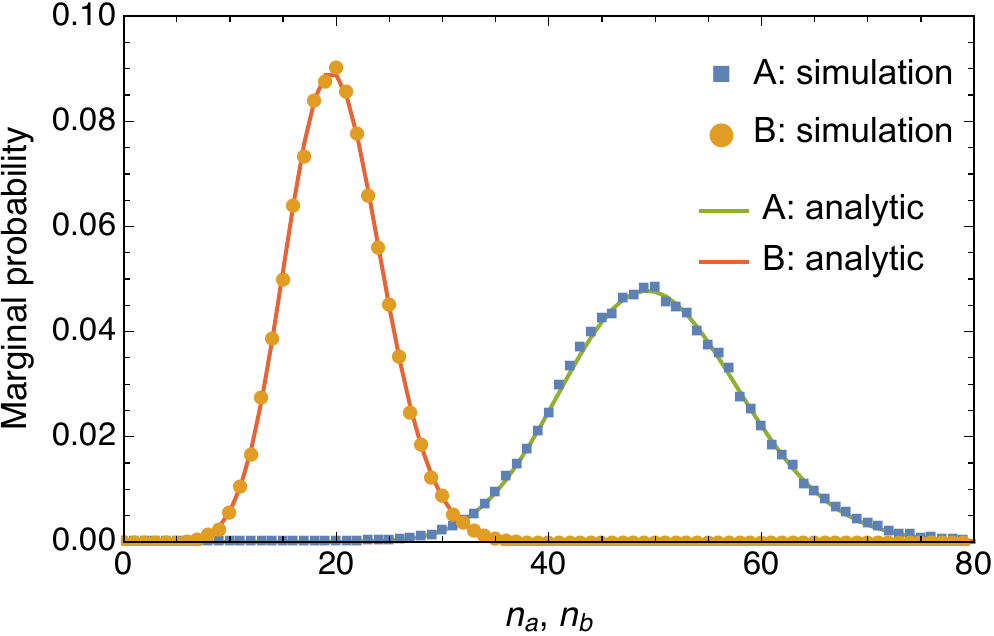} 
\caption{ 
Marginal probability distributions of $n_a$ and $n_b$ from the analytic solution and numerical simulations. 
Solid lines are from the analytic solution,
which are consistent with the result of Gillespie simulations indicated by boxes (species A) and circles (species B). 
Parameters are chosen as $k_1=10,k_2=1,k_3=20,k_4=1,k_5=20$,
and we have simulated $50,000$ events for the plot. 
}
\label{fig:eg-nl-marginal}
\end{figure}
The normalized solution is given by 
\begin{equation}
\Psi' (z_a,z_b) 
=
\exp 
\left[ 
\frac{k_5}{2k_2} ((z_a)^2-1) + 
\frac{k_1 + k_5}{k_2} (z_a - 1) 
+ \frac{k_3}{k_4} (z_b-1) 
\right]. 
\end{equation}
The stationary distribution of $\Gamma'$ 
is analytically obtained as 
\begin{equation}
P_{\rm s} (n_a,n_b) 
= 
e^{
- \frac{1}{k_2}
\left( \frac{3}{2} k_5 + k_1 \right) 
- 
\frac{k_3}{k_4} 
}
\frac{1}{n_a !}
\left( 
- \frac{k_5}{2k_2 } 
\right)^{\frac{n_a}2}
H_{n_a}
\left( \frac{k_1+k_5}{\sqrt{ -2 k_2 k_5}} \right) 
\frac{1}{n_b!} 
\left(\frac{k_3}{k_4}\right)^{n_b} .
\label{eq:eg-nl-prob-analytic}
\end{equation}
We have checked that this expression is consistent with numerical simulations based on the Gillespie algorithm (Fig.~\ref{fig:eg-nl-marginal}).

\section{ Squeezing generic chemical reaction systems }\label{sec:generic}

So far, we have discussed squeezing transformations in three simple examples. 
One can perform squeezing on more complicated reaction networks. 
For example, the single-mode squeezing of
species $A$ on the following network results in: 
\begin{equation}
\begin{tikzpicture} 

\node at (-1, 0) { 
{ 
$
\begin{tikzcd}
A
\rar["\ell_1", shift left] 
&
B
\lar["\ell_2", shift left]
& 
\emptyset 
\rar["\ell_3", shift left] 
& 
A+B
\lar["\ell_4", shift left]
\end{tikzcd} 
$
} 
};  

\node at (9, -0.3) { 
{ 
$
\begin{tikzcd}
A
\rar["k_1", shift left] 
&
B
\lar["k_2", shift left]
\dar["k_6", shift left]
& 
\emptyset 
\rar["k_3", shift left]
\dar["k_5", shift left] 
& 
A+B
\lar["k_4", shift left]
\\
& 2A + B & 2A
&
\end{tikzcd} 
$
} 
};  

\draw[mybrightblue, -stealth, line width=4pt] 
(2.5,0) -- (4.8,0); 

\node at (3.6,-0.6){ \scalebox{1} {\color{mybrightblue} Squeezing }};





\end{tikzpicture}     
\label{eq:squeeze-tr-eg}
\end{equation}
It is natural to ask 
what kind of structural transformation 
is induced in squeezing for a 
generic reaction network, 
which we discuss here. 
Recall that the stochastic Hamiltonian 
for a generic reaction network is 
written as 
\begin{equation}
H = 
\sum_A k_A \left[ 
(a^\dag)^{t_A} - (a^\dag)^{s_A} 
\right] a^{s_A} . 
\end{equation}
We shall pick one species $v_i$ 
and act the two-mode squeezing operator 
mixing $v_i$ and $v_j$ (the case of single-mode squeezing can be obtained by setting $j=i$). 
The annihilation operator $a_i$ is transformed as 
\begin{equation}
a_i \mapsto S_j (r) a_i S_j^{-1} (r) 
= \cosh r \, a_i + \sinh r \, a^\dag_j ,
\end{equation}
For simplicity, we here assume that species $v_i$ appears only once as a source in a reaction 
(namely, $s_{iA} \le 1$ for any $A$). 
As a transformed Hamiltonian, 
we consider the following: 
\begin{equation}
H' =     
\sum_A 
k_A \left[ 
(a^\dag)^{t_A} - 
(a^\dag)^{s_A} 
\right] 
S_j (r) a^{s_A} S^{-1}_j (r) .
\end{equation}
If the reaction $e_A$ contains one $v_i$ as a reactant, 
\begin{equation}
S_j (r) a^{s_A} S_j^{-1} (r)
=
(\cosh r\, a_i + \sinh r \, a_j^\dag) 
a^{s_A - e_{i}} ,
\end{equation}
where $e_{i} \in \mathbb N^V$ is a vector 
whose $i$-th component is one 
and other components are zero 
if $i$ is a source of reaction $A$
and otherwise a zero vector. 
This Hamiltonian $H'$ has a steady state that satisfies $H'\ket{\Psi'}=0$ given by a displaced squeezed state,
\begin{eqnarray}
\ket{\Psi'} = S_i(r) \ket{\Psi}. 
\end{eqnarray}
In the transformed Hamiltonian, 
the part containing one $v_i$ 
as a source is written as 
%
%
%
%
\begin{equation}
\begin{split}
&
k_A \left[ 
(a^\dag)^{t_A} - 
(a^\dag)^{s_A} 
\right] 
(\cosh r\, a_i + \sinh r\,  a_j^\dag)
a^{s_A - e_{i}} 
\\
&= 
k_A \cosh r
\left[ 
(a^\dag)^{t_A} - 
(a^\dag)^{s_A} 
\right] 
a^{s_A} 
+ 
k_A \sinh r 
\left[ 
(a^\dag)^{t_A + e_{j} } 
- (a^\dag)^{s_A + e_{j} } 
\right] 
a^{s_A - e_{i}}
.
\end{split}
\end{equation}
We write the second term as 
\begin{equation}
\begin{split}
\text{(2nd term)}
&= 
k_A \sinh r 
\left[ 
(a^\dag)^{t_A + e_{j}} - (a^\dag)^{s_A + e_{j}} 
\right] 
a^{s_A - e_{i}} 
\color{mydarkred}
+
k_A \sinh r
(a^\dag)^{s_A - e_{i}}
a^{s_A - e_{i}} 
\color{mydarkpurple}
- 
k_A \sinh r
(a^\dag)^{s_A - e_{i}}
a^{s_A - e_{i}} 
\color{black}
\\
&= 
k_A \sinh r
\left[ 
(a^\dag)^{t_A + e_{j}} - (a^\dag)^{s_A - e_{i}} 
\right] 
a^{s_A - e_{i}} 
- 
k_A \sinh r
\left[ 
(a^\dag)^{s_A + e_{j}} - (a^\dag)^{s_A - e_{i}} 
\right] 
a^{s_A - e_{i}} ,
\end{split}
\label{eq:2nd-term}
\end{equation}
%
%
where the colored part sums up to zero. 
From Eq.~\eqref{eq:2nd-term}, 
we can see that two additional reactions appear, 
whose chemical content can be read off. 
Namely, if the reaction $e_A$ contains 
one $v_i$ as a reactant, 
the transformed reaction system 
contains the following two 
additional reactions\footnote{
Similarly, if there is a reaction containing one $v_j$ as a source, two additional reactions appear from the reaction via the two-mode squeezing. 
}, 
\begin{align}
s_A - e_{i} 
\quad &\longrightarrow 
\quad s_A + e_{j}, 
\label{eq:add-reac-1}
\\
s_A - e_{i} 
\quad 
&\longrightarrow 
\quad 
t_A + e_{j}. 
\label{eq:add-reac-2}
\end{align}
The reaction constants of 
these reactions are given by 
$k_A \sinh r$ 
and $-k_A \sinh r$. 
If the original reaction contains 
either of the two additional reactions, 
it is possible to take all the reaction coefficients positive in the resulting system. 
If we set $j=i$, we obtain the additional reactions in the case of single-mode squeezing of species $v_i$.

One can check that all the examples discussed 
earlier can be understood from the rules ~\eqref{eq:add-reac-1} and 
\eqref{eq:add-reac-2}. 
For example, for the example~\eqref{eq:squeeze-tr-eg}, 
there are two reactions that involve $A$ as its source. 
From the reaction $A \to B$, there appear the following two reactions, 
\begin{align}
    \emptyset &\longrightarrow 2A, \\
    \emptyset &\longrightarrow A+B.
\end{align}
The second one already exists in the original set of reactions,
and the transformation renormalizes the rate constant of the reaction. 
From the reaction $A +B \to \emptyset$, we have 
\begin{align}
    B &\longrightarrow 2A+B, \\
    B &\longrightarrow A,
\end{align}
where the second one is in the original reactions. 
As a result, we obtain the network shown on the right of \eqref{eq:squeeze-tr-eg}.

\section{ Summary and discussion }\label{sec:summary}

In this paper, we studied the stationary distributions of stochastic chemical reaction systems using an analogy to the quantum mechanics. 
Stationary distributions of the product Poisson form 
in the Anderson--Craciun--Kurtz theorem correspond to coherent states in the quantum-mechanical formulation, and we considered squeezing of the coherent states. 
Using the same squeeze operator, 
the stochastic Hamiltonian is also transformed, 
and the squeezed coherent states are the zero eigenstate of the new Hamiltonian. 
The transformed Hamiltonian represents a different chemical reaction network from the original one, 
and in general its deficiency is nonzero and 
weak reversibility is lost. 
From the squeezed coherent states, we can obtain analytical expressions of the stationary distribution of the new reaction network. 
We validated the obtained expressions of stationary distributions via comparison with stochastic simulations. 
We also discussed the form of additional reactions that appear in a squeezing for a generic chemical reaction network.

The present method can be applied even when the reaction network is higher than first order and certain reactions involve two or more species as reactants. 
However, we admit that the reaction networks that can be reached by squeezing those with complex-balanced equilibria are rather limited.
Still, we believe that it would be possible to find analytic stationary distributions by considering other kinds of transformation\footnote{
A duality relation for stochastic processes has been discussed~\cite{Ohkubo_2013} based on the Doi--Peliti formalism. 
}, which may or may not come from the analogy to quantum mechanics. 
In this paper, we considered single-mode and two-mode squeezed states. 
There are other types of multiphoton coherent states~\cite{DELLANNO200653} 
and it would be interesting to examine their counterparts in stochastic chemical systems. 

\begin{acknowledgments}
The authors are grateful to Hyukpyo Hong and Bryan Hernandez for helpful discussions. 
Y.~H. and R.~H. are supported by an appointment of the JRG Program at the APCTP, which is funded through the Science and Technology Promotion Fund and Lottery Fund of the Korean
Government, and is also supported by the Korean Local
Governments of Gyeongsangbuk-do Province and Pohang City. 
Y.~H. is also supported by the National Research Foundation (NRF) of Korea (Grant No. 2020R1F1A1076267) funded by the Korean Government (MSIT).
\end{acknowledgments}

\appendix

\bibliography{refs}

\end{document}